\documentclass[aps,twocolumn,superscriptaddress,amsmath,amssymb,showpacs]{revtex4-2}
\usepackage{graphicx}
\usepackage{dcolumn}
\usepackage[absolute,overlay]{textpos}

\usepackage{bm}
\usepackage[colorlinks, allcolors=blue]{hyperref}
\usepackage[usenames, dvipsnames]{color}
\usepackage{setspace}
\usepackage{scalerel}
\usepackage[utf8]{inputenc}
\usepackage[T1]{fontenc}
\usepackage{hyperref}
\usepackage{natbib}
\usepackage[normalem]{ulem}
\usepackage{cancel}
\usepackage{amsmath}
\usepackage{amsfonts}
\usepackage{stackengine}
\usepackage{comment}
\usepackage[scr=esstix,cal=boondox]{mathalfa}
\usepackage{mathrsfs}
\usepackage{csquotes}
\usepackage{soul}
\newcommand{\ag}{\textcolor{black}}
\stackMath
\newsavebox\tmpbox
\begin{document}

\title{Identifying the net information flow direction pattern in mutually coupled non-identical chaotic oscillators}
\author{Anupam Ghosh}
\email{ghosh@cs.cas.cz}
\address{Department of Complex Systems, Institute of Computer Science of the Czech Academy of Sciences, Prague 18200, Czechia}
\author{X. San Liang} 
\address{Division of Frontier Research, Southern Marine Science and Engineering Guangdong Laboratory, Zhuhai 519000, China}
\address{Department of Atmospheric \& Oceanic Sciences, Fudan University, Shanghai 200438, China}
\author{Pouya Manshour}
\address{Department of Complex Systems, Institute of Computer Science of the Czech Academy of Sciences, Prague 18200, Czechia}
\author{Milan Palu\v{s}} 
\address{Department of Complex Systems, Institute of Computer Science of the Czech Academy of Sciences, Prague 18200, Czechia}
\begin{abstract}
This paper focuses on a fundamental inquiry in a coupled oscillator model framework. It specifically addresses the direction of net information flow in mutually coupled non-identical chaotic oscillators. Adopting a specific form of conditional mutual information as a model-free and asymmetric index, we establish that if the magnitude of the maximum Lyapunov exponent can be defined as the `degree of chaos' of a given isolated chaotic system, a predominant net information transfer exists from the oscillator exhibiting a higher degree of chaos to the other while they are coupled. We incorporate two distinct categories of coupled `non-identical' oscillators to strengthen our claim. In the first category, both oscillators share identical functional forms, differing solely in one parameter value. We also adopt another measure, the Liang--Kleeman information flow, to support the generality of our results. The functional forms of the interacting oscillators are entirely different in the second category. We further extend our study to the coupled oscillator models, where the interacting oscillators possess different dimensions in phase space. These comprehensive analyses support the broad applicability of our results.
\end{abstract}
\maketitle
\section{Introduction}
\label{sec:intro}
A complex system consists of many interacting elements, and its complex structure and nonlinear interactions among subsystems cause the emergence of many nonlinear phenomena, such as synchronization, oscillation quenching, and swarming~\cite{winfree01,pikovsky01,lakshmanan03,balanov08}. Thus, the dynamics of a complex system cannot be inferred by studying the subsystems separately, and coupling between the subsystems plays a significant role in determining the complete dynamical state. A coupled oscillator model (COM) is useful for studying the aforementioned nonlinear phenomena. These models consist of multiple oscillators, which interact through a coupling mechanism that can be either linear or nonlinear. COMs have widely been used to investigate the collective behavior of interacting systems, ranging from physical systems like atoms and molecules to biological systems like cells and neurons~\cite{pikovsky01,balanov08}. Consequently, these models have found applications in various fields, such as physics, chemistry, biology, and engineering, for understanding and predicting the behavior of complex systems~\cite{winfree01,pikovsky01,lakshmanan03,balanov08}.
This paper focuses on investigating the direction of coupling in a COM of two non-identical chaotic oscillators. Identifying the driver and driven oscillators for unidirectional coupling is straightforward, i.e., information is transferred from the driving oscillator to the driven oscillator. This notion of driver--driven relationship is inapplicable in the case of mutual (or bidirectional) coupling. Instead, we can ask the fundamental question: \emph{What is the direction of net information transfer in mutually coupled non-identical oscillator models?} Being more explicit, we are interested in whether the first oscillator drives the second oscillator more or vice-versa. On the same track, another related question we can ask: \emph{Is there any pattern of this net information transfer direction?}
Knowing the direction of interaction between the subsystems is the key to understanding this net information transfer. In this paper, we use a particular form of conditional mutual information (CMI), proposed by Palu\v{s} et al.~\cite{palus01}, as the required index to identify the direction of the interaction. CMI is a statistical measure used to quantify the dependence between two variables while considering the influence of a third variable. It provides a way to determine how much the first variable can be predicted by knowing the second variable, given the information provided by a third variable. Note that, unlike mutual information~\cite{cover06,ghosh22,ghosh22_2}, this specific form of CMI is an asymmetric measure. This index CMI is the generalization of the notion of Granger causality~\cite{granger69} for a Gaussian process. A later study analytically proves the equivalence between the transfer entropy~\cite{schreiber00} and the CMI for a Gaussian process~\cite{barnett09}. Thus, a directed coupling can be interpreted as the transfer of information from one system to the other~\cite{gupta21}. Subsequent studies~\cite{runge12,kugiumtzis13,sun14,manshour21,manshour24} have extended the applicability of these indices from bivariate (interaction between two processes) to multivariate case (interaction among multiple processes). However, in this paper, we restrict ourselves to the bivariate case. The cross-scale interactions of atmospheric dynamics of recorded daily surface temperature from different European locations have been studied using this measure~\cite{palus14}. The detection of causality in simulated data of the epileptor model and scrutinizing the information transfer using CMI have also been done~\cite{gupta21}. CMI has been used to detect cross-frequency causality in simulated epileptic seizures, as well as in real EEG data from an animal experiment~\cite{palus24}. Recently, CMI has also been utilized to identify the cause responsible for the increased probability of cold extremes in the winter and spring surface air temperature in Europe, demonstrating the ability of CMI to detect the cause of extreme events~\cite{palus24}. Therefore, CMI is a powerful tool for analyzing complex systems and has numerous applications, including estimating information flow in neural networks, data compression, and clustering~\cite{runge12,kugiumtzis13,sun14, palus14,gupta21}.
In order to support the results associated with the aforementioned approach for assessing directionality, this study incorporates another novel and tractable technique called information flow, or the Liang--Kleeman (LK) information flow, Liang information flow, etc., as referred to in the literature ~\cite{liang05,liang16}. This method, also grounded in information theory, represents a rigorous formalism of information transfer between dynamical events~\cite{liang05,liang14,liang15}. The LK information flow technique can be used to evaluate, in a rigorous sense, the cause and effect in dynamical systems and has been successfully applied to numerous problems across various disciplines, such as brain dynamics~\cite{hristopulos19}, climate research~\cite{liang16}, and quantum mechanics~\cite{yi22}.
Along with CMI and LK information flow, other measures are also reported in the literature to detect the directionality of the coupling. Arnhold et al.~\cite{arnhold99} have reported an asymmetric measure based on the distances of points in state space. Another directionality index~\cite{rosenblum01,ghosh_23_conf} is reported, which can measure the cross-dependence of phases in a COM. This method is applicable to chaotic and noise-perturbed systems. The calculation of correlation dimension is an important measure to study the causal relation~\cite{janjarasjitt08,krakovska19}. The permutation CMI~\cite{li10} and symbolic transfer entropy~\cite{staniek08} are also used in the literature to explore causality. However, the symbolic transfer entropy approach does not incorporate the amplitude part of the original data and only analyzes the phase part; as a remedy, another index, termed weighted symbolic transfer entropy~\cite{li20}, has recently been reported. The Compression-Complexity Causality is also a model-free measure~\cite{kathpalia19} that overcomes the limitations of Granger causality~\cite{granger69} and transfer entropy~\cite{schreiber00} up to a certain extent. These methods help study the coupling direction among the neural oscillators, which further helps to identify the information flow in different regions of the brain.    
In this study, we utilize the CMI and the LK information flow as the required indices to investigate non-identical COMs. Our results show that in all cases, the oscillator with a higher maximum Lyapunov exponent value, calculated when the coupling is inactive, transfers more information to the oscillator with a lower maximum Lyapunov exponent value. This indicates that \emph{a predominant net information transfer direction exists from the oscillator with a higher maximum Lyapunov exponent to the other}. The maximum Lyapunov exponent~\cite{strogatz07} is a measure of the sensitivity of the system's behaviour to initial conditions in phase space and provides information about the predictability of the system's evolution. A positive value of the maximum Lyapunov exponent indicates that the system is chaotic and slight differences in the initial conditions can lead to significant differences in the system's future behaviour.
This paper is organized as follows: Section~\ref{sec:cmi} provides a detailed discussion on the index CMI. In Sec.~\ref{sec:lk_index}, we explore the LK information flow. Section~\ref{sec:para_mis} presents the calculations of these two directionality indices using non-identical R\"ossler oscillators~\cite{roessler76}. In Secs.~\ref{sec:non_iden} and \ref{sec:diff_topo}, we compute the CMI values for various examples of non-identical COM. Finally, Sec.~\ref{sec:conclusion} concludes with a summary of the main results of this paper.
\section{Results and discussion}
\label{sec:result}
The notion of coupled `non-identical' oscillators can be categorized into two classes: (a) The functional forms of both oscillators are identical, with a mismatch in one (or more) parameter value(s). These types of oscillators are also termed nearly-identical oscillators. (b) The corresponding functional forms are entirely different for both oscillators. We have incorporated both classes of COMs in this paper. However, before the detailed analysis, we discuss CMI and LK information flow more elaborately in Secs.~\ref{sec:cmi} and \ref{sec:lk_index}, respectively.    
\subsection{Conditional Mutual Information (CMI)}
\label{sec:cmi}

Let $X = \{ x_j\}_{j = 1}^{N}$, $Y = \{ y_j\}_{j = 1}^{N}$, and $Z = \{ z_j\}_{j = 1}^{N}$ be three discrete variables. The Shannon entropy~\cite{cover06} of the variable $X$ is given by: 
\begin{equation}
	\label{eq:shanon}
	H(X) = - \sum_{j=1}^{N} p(x_j) \log_{2} \left(p(x_j) \right),
\end{equation}
where $p(x_1), p(x_2), \cdots, p(x_N)$ are the probabilities associated with $x_1, x_2, \cdots, x_N$, respectively. The base $2$ in the logarithm function of Eq.~\ref{eq:shanon} implies that $H(X)$ comes in the unit of bits. Intuitively, $H(X)$ quantifies the amount of uncertainty that the variable $X$ contains~\cite{cover06}. Likewise, we can define the Shannon entropy $H(Y)$ and  $H(Z)$ for variables $Y$ and $Z$, respectively.
Furthermore, we can define the joint entropy $H(X,Y)$ of variables $X$ and $Y$:
\begin{equation}
	\label{eq:joint_entropy}
	H(X, Y) = - \sum_{j,k=1}^{N} p(x_j, y_k) \log_{2} \left(p(x_j, y_k) \right).
\end{equation}
Similarly, the conditional entropy $H(X|Y)$ of $X$ when $Y$ is given is as followed:
\begin{equation}
	\label{eq:condi_entropy}
	H(X| Y) = - \sum_{j,k=1}^{N} p(x_j, y_k) \log_{2} \left(\frac{p(x_j, y_k)}{p(y_k)} \right), 
\end{equation}
with $p(y_k) \neq 0$. The CMI $I(X;Y|Z)$ of $X$ and $Y$ when $Z$ is given is as follows:
\begin{equation}
	\label{eq:cmi_1}
	I(X;Y|Z) = H(X|Z) + H(Y|Z) - H(X,Y|Z).
\end{equation}
If $X$ and $Y$ are independent of $Z$, we can write
\begin{equation}
	I(X;Y|Z) = I(X;Y),
\end{equation}
where $I(X;Y) = H(X) + H(Y) - H(X,Y)$ is the mutual information between the variables $X$ and $Y$. After a simple manipulation, Eq.~\ref{eq:cmi_1} reduces to  
\begin{equation}
	\label{eq:cmi_2}
	I(X;Y|Z) = I(X;Y;Z) -I(X;Z) - I(Y;Z),
\end{equation}
where $I(X;Y;Z) = H(X) + H(Y) + H(Z) - H(X,Y,Z)$. The value of CMI can vary within the range between $0$ and $\text{max}\{H(X), H(Y)\}$~\cite{vlachos22}.

In this paper, we calculate the CMI value between two variables obtained from mathematical models of interacting oscillators. For any two such variables $\{ x''_j\}_{j = 1}^{N}$ and $\{ x'_j\}_{j = 1}^{N}$, the amount of information transfer from $x''_j$ to the future state $x'_{j+\tau}$ is given by
\begin{widetext}
\begin{equation}
	\label{eq:cmi_delay1}
	I(X;Y|Z) = I\left(\{x''_j, x''_{j-\rho_0}, \cdots, x''_{j-(d_1-1)\rho_0}\}; x'_{j+\tau}|\{x'_j, x'_{j-\eta_0}, \cdots, x'_{j -(d_2-1)\eta_0} \}\right),
\end{equation}
\end{widetext}
with 
\begin{subequations}
	\begin{eqnarray}
		X &=& \{x''_j, x''_{j-\rho_0}, \cdots, x''_{j-(d_1-1)\rho_0}\},\\
		Y &=& x'_{j+\tau},\\
		Z &=& \{x'_j, x'_{j-\eta_0}, \cdots, x'_{j -(d_2-1)\eta_0} \},
	\end{eqnarray}
\end{subequations}
where $\rho_0$ and $\eta_0$ are the embedding delays for the variables $\{ x''_j\}_{j = 1}^{N}$ and $\{ x'_j\}_{j = 1}^{N}$, respectively. These embedding delays are generally calculated using the method proposed by Fraser and Swinney~\cite{fraser86}. Besides, $d_1$ and $d_2$ are the respective minimum embedding dimensions. An earlier study~\cite{palus07} suggests that the calculation of CMI using variables $X = x''_j$, $Y = x'_{j + \tau}$, and $Z = \{x'_j, x'_{j-\eta_0}, x'_{j -2\eta_0} \}$ is sufficient. Therefore, Eq.~\ref{eq:cmi_delay1} reduces to  
\begin{equation}
	\label{eq:cmi_delay2}
	I(X;Y|Z) = I\left(x''_j; x'_{j+\tau}|\{x'_j, x'_{j-\eta_0}, x'_{j -2\eta_0} \}\right).
\end{equation}
Note that while mutual information $I(X;Y)$ measures the mutual dependence between $X$ and $Y$, CMI $I(X;Y|Z)$ returns the direction of coupling from $X$ to $Y$ without the possible influence of $Z$.
\subsection{Liang--Kleeman (LK) information flow}
\label{sec:lk_index}
Here, we briefly overview the second index, the LK information flow, which we use in this study. In $2005$, Liang and Kleeman~\cite{liang05} discovered a law in dynamical systems, which leads to the establishment of a rigorous formalism of information flow later on~\cite{liang08}; see Ref.~\cite{liang16} for a historic document. In order to understand the mechanism of this index, consider the following 2D coupled stochastic system, as originally considered by Liang~\cite{liang08}:
\begin{equation}
	\label{eq:stochastic}
	d\mathbf{X} = \mathbf{F}(\mathbf{X}, t) \cdot dt + \mathbf{B}(\mathbf{X}, t) \cdot d\mathbf{W},
\end{equation}
where $\mathbf{X} = (x_1, x_2)$ is the vector of variables, the vector field $\mathbf{F}=(F_1, F_2)$ contains the drift coefficients and it is differentiable, $\mathbf{B} = (b_{ij})$ is the stochastic perturbation coefficients matrix, and $\mathbf{W}=(W_1, W_2)$ is a 2D standard Wiener process. Let $g_{ij} = \sum_{k=1}^{2} b_{ik} b_{jk}$, and $\rho_i$ denotes the marginal probability density function of $x_i$. It has been proven~\cite{liang08} that the time rate of information flowing from $x_2$ to $x_1$ is
\begin{equation}
	\label{eq:t21}
	T_{x_2 \rightarrow x_1} = - E\left( \frac{1}{\rho_1} \frac{\partial (F_1 \rho_1)}{\partial x_1}\right) + \frac{1}{2}E\left( \frac{1}{\rho_1} \frac{\partial^2 g_{11} \rho_1}{\partial x^2_1}\right).
\end{equation}
Here, $E$ represents the expectation value. The index $T_{x_2 \rightarrow x_1}$ measures the asymmentric information flow between the variables $x_1$ and $x_2$. The expression for $T_{x_2 \rightarrow x_1}$ is rigorous; however, calculating the information flow in real-world problems necessitates previous knowledge of Eq.~\ref{eq:stochastic}, which is typically unknown. In order to address this, Liang~\cite{liang14} made a maximum likelihood of Eq.~\ref{eq:t21} with a linear model. This way, the LK information flow ($T_{x_2 \rightarrow x_1}$) can be estimated as follows:
\begin{equation}
	\label{eq:t21_final}
	T_{x_2 \rightarrow x_1} =   \frac{C_{11}C_{12} C_{2,d1} - C^2_{12} C_{1,d1}}{C^2_{11} C_{22} - C_{11} C^2_{12}},
\end{equation}
where $C_{ij}$ denotes the sample covariance between $x_i$ and $x_j$, and $C_{i,dj}$ is the aforesaid covariance between $x_i$ and the time derivative of $x_j$. Note that if the variable $x_1$ is independent of $x_2$, the information flow from $x_2$ to $x_1$ is zero, indicating that $x_2$ is not causal to $x_1$. When $T_{x_2 \rightarrow x_1}$ is nonzero, it can take either positive or negative values. A positive value of $T_{x_2 \rightarrow x_1}$ implies that $x_2$ increases the uncertainty of $x_1$, whereas a negative value of $T_{x_2 \rightarrow x_1}$ reduces the entropy of $x_1$~\cite{liang14}.
\subsection{COM with parameter mismatch}
\label{sec:para_mis}
\begin{figure}[h]
	\centering
	\includegraphics[width=30cm,height=2.9cm, keepaspectratio]{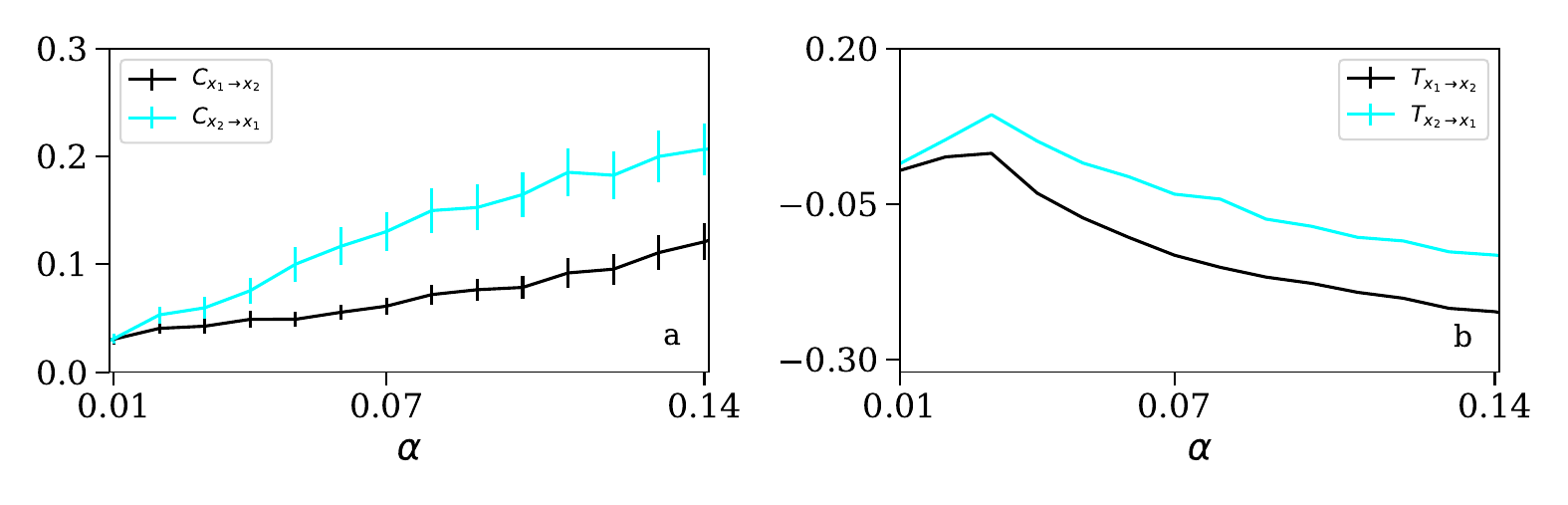}
	\caption{The CMI values and the LK information flow indices have been calculated for coupled R\"ossler oscillators (Eq.~\ref{eq:ross}) and plotted as a function of $\alpha$ in subplots~(a) and (b), respectively. In subplot~(a), CMI values become distinguishable for higher $\alpha$ values. Similarly, $T_{x_1 \rightarrow x_2}$ and $T_{x_2 \rightarrow x_1}$ are dissimilar over the same range of $\alpha$. The error bars in subplot~(b) are not visible because of their smaller values ($\sim 10^{-4}$).} 
	\label{fig:ross2}
\end{figure}

The general form of equations of motion is as follows:
\begin{subequations}
	\label{eq:para}
	\begin{eqnarray}
		\frac{d{\mathbf{x}}_1}{dt} &=& \mathbf{F(x}_1, \mu_1) + \alpha \cdot \mathsf{C} \cdot (\mathbf{x}_2 - \mathbf{x}_1),\label{eq:para_a}\\
		\frac{d{\mathbf{x}}_2}{dt} & =& \mathbf{F(x}_2, \mu_2) + \alpha \cdot \mathsf{C} \cdot  (\mathbf{x}_1 - \mathbf{x}_2),\label{eq:para_b}
	\end{eqnarray}
\end{subequations}
where vectors $\mathbf{x}_1$ and $\mathbf{x}_2$ are the phase space coordinates of the respective oscillators; both oscillators have the same functional form, $\mathbf{F}(\cdot)$. The different values of parameters $\mu_1$ and $\mu_2$ make these two oscillators non-identical. Besides, $\alpha$ is the coupling strength and $\mathsf{C}$ is the coupling matrix. Since we mostly deal with three-dimensional oscillators in this paper, the coupling matrix $\mathsf{C}$ has the dimension $3 \times 3$ (except for Sec.~\ref{sec:diff_topo}). To this end, we mention that in this paper, the fourth-order Runge--Kutta method has been used with the smallest time interval of $0.01$ and the maximum time of $1000$ to solve the ordinary differential equations numerically. We remove initial $5\%$ data as transient and use the rest $95\%$ data for analysis. 
Now, we adopt the first example of this paper: mutually coupled non-identical R\"ossler~\cite{roessler76} oscillators. It is the example of an autonomous, three-dimensional oscillator, which shows chaotic dynamics at some specific combinations of its parameter values. The explicit form of coupled R\"ossler oscillators is given by:
\begin{subequations}
	\label{eq:ross}
	\begin{eqnarray}
		\frac{d{x}_m}{dt} &=& -y_m-z_m + \alpha (x_n - x_m),\label{eq:ross_a}\\
		\frac{d{y}_m}{dt} &=& x_m+ 0.2 y_m,\label{eq:ross_b}\\
		\frac{d{z}_m}{dt} &=& 0.2 + z_m (x_m - c_m),\label{eq:ross_c}
	\end{eqnarray}
\end{subequations}
where $m = 1,2$ and $n = 1,2$ with $n \neq m$. The system parameters are $c_1 = 5.7$ and $c_2 = 10$. For the example in hand, following Eq.~\ref{eq:para}, the coupling matrix $\mathsf{C}$ has one non-zero element, i.e., $\mathsf{C} = \text{diag}(1, 0, 0)$. We can also write $\mathbf{x}_1 = (x_1, y_1, z_1)$ and $\mathbf{x}_2 = (x_2, y_2, z_2)$.
We calculate the CMI using variables $x_1 (t)$ and $x_2(t)$ at different values of $\alpha$ using Eqs.~\ref{eq:cmi_delay2} and \ref{eq:ross}. With the analogy of $\{ x''_j\}_{j = 1}^{N} = x_1(t)$ and $\{ x'_j\}_{j = 1}^{N} = x_2(t)$, Eq.~\ref{eq:cmi_delay2} measures the amount of information transfer from $x_1(t)$ to $x_2(t+\tau)$ and abbreviates it as $C_{x_1 \rightarrow x_2}$. Likewise, we can define $C_{x_2 \rightarrow x_1}$ to measure the amount of information transfer from $x_2(t)$ to $x_1(t+\tau)$. \ag{A well-established technique~\cite{fraser86} for determining the appropriate embedding delay involves computing the mutual information between two temporally spaced data points, $x'_j$ and $x'_{j+\eta}$, within the given time series $\{ x'_j\}_{j = 1}^{N}$. Subsequently, the corresponding embedding delay, denoted as $\eta = \eta_0$, is selected at which the mutual information function is around its first minimum. For coupled R\"ossler oscillators, following the aforementioned technique, we adopt $\eta_0 = 180$ for both variables $x_1 (t)$ and $x_2(t)$.} Similarly, $\eta_0 = 20, 15,$ and $200$ for Lorenz~\cite{lorenz63}, Chen~\cite{chen99}, and 4D R\"ossler oscillators~\cite{ross79}, respectively. The aforementioned values of $\eta_0$ will be used in Secs.~\ref{sec:non_iden} and \ref{sec:diff_topo} to calculate the CMI values. Although, in this paper, we choose the $x$-coordinates of coupled oscillators to calculate CMIs, one may repeat these analyses using the $y$-coordinates, eventually leading to similar conclusions.
\begin{figure}[h]
	\centering
	\includegraphics[width=30cm,height=2.9cm, keepaspectratio]{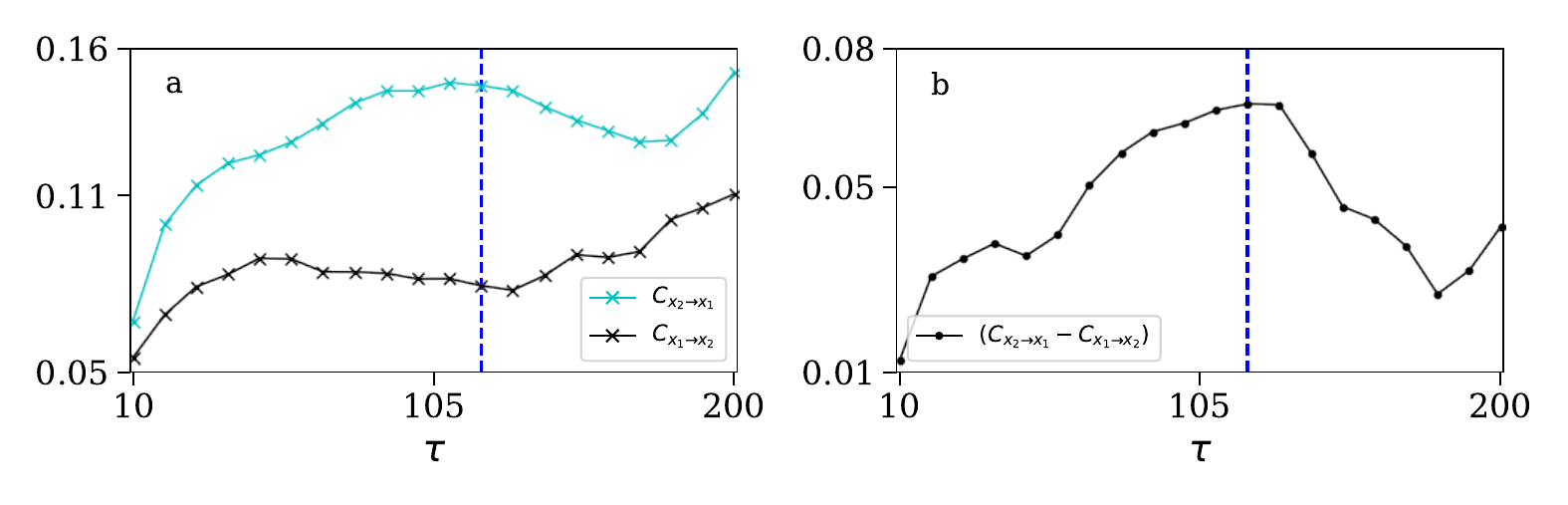}
	\caption{The CMI values have been calculated for coupled R\"ossler oscillators (Eq.~\ref{eq:ross}) and plotted as a function of $\tau$ for a fixed $\alpha = 0.15$. In subplot~(a), the cyan plot, $C_{x_2 \rightarrow x_1}$, always has higher values than the black plot ($C_{x_1 \rightarrow x_2}$). The difference between $C_{x_2 \rightarrow x_1}$ and $C_{x_1 \rightarrow x_2}$ is depicted as a function of $\tau$ in subplot~(b). The vertical blue dashed lines in both subplots correspond to $\tau = 120$.} 
	\label{fig:ross1}
\end{figure}
Figure~\ref{fig:ross1}a depicts the variation of the CMI values as a function of time delay ($\tau$) at an arbitrary coupling strength $\alpha = 0.15$. The index $C_{x_1 \rightarrow x_2}$ returns the influence of  $x_1(t)$ on $x_2(t+\tau)$, whereas $C_{x_2 \rightarrow x_1}$ measures the influence of $x_{2} (t)$ on $x_{1} (t + \tau)$. It is clearly visible that $C_{x_2 \rightarrow x_1}$ (cyan plot) always has higher values than $C_{x_1 \rightarrow x_2}$ (black plot). This observation further creates curiosity to study the variations of indices $C_{x_1 \rightarrow x_2}$ and $C_{x_2 \rightarrow x_1}$ as a function of $\alpha$. In order to do so, one has to adopt an appropriate $\tau$, at which the difference between $C_{x_2 \rightarrow x_1}$ and $C_{x_1 \rightarrow x_2}$ is significant. Figure~\ref{fig:ross1}b supports that the difference between $C_{x_2 \rightarrow x_1}$ and $C_{x_1 \rightarrow x_2}$ is maximum at $\tau = 120$. This maximum position varies along the $\tau$-axis for different values of $\alpha$. In principle, any value of $\tau$ should work; however, sometimes, the differences between $C_{x_1 \rightarrow x_2}$ and $C_{x_2 \rightarrow x_1}$ are negligible for smaller $\tau$, as shown in Fig.~\ref{fig:ross1}b. Finally, we compute the average CMI values across $400$ different $\tau$-values to minimize the influence of $\tau$ dependence on the outcome. The calculated average CMI values at different values of $\alpha$ are depicted in Fig.~\ref{fig:ross2}a. The height of the error bars is nothing but the standard deviation of the $400$ different CMI values. As we progress from $\alpha = 0.01$ onward, a distinct shift transpires: $C_{x_2 \rightarrow x_1}$ begins to display higher values than $C_{x_1 \rightarrow x_2}$. Figure~\ref{fig:ross2}b illustrates the variation of the LK information flow indices as a function of the coupling strength ($\alpha$). Here also, to minimize the influence of $\tau$ dependence on the outcome, we compute the average the LK information flow indices across $400$ different $\tau$ values. The index $T_{x_1 \rightarrow x_2}$ represents the influence of $x_1(t)$ on $x_2(t+\tau)$, whereas $T_{x_2 \rightarrow x_1}$ measures the influence of $x_2(t)$ on $x_1(t+\tau)$. It is evident that $T_{x_2 \rightarrow x_1}$ (cyan plot) consistently exhibits higher values than $T_{x_1 \rightarrow x_2}$ (black plot). Thus, the parameter mismatch in Eq.~\ref{eq:ross} implies a net information transfer from one oscillator to the other. 
Furthermore, we have calculated the maximum Lyapunov exponent ($\lambda_{\rm max}$)~\cite{sprott03} for Eq.~\ref{eq:ross} with $\alpha = 0$. The first oscillator has $\lambda_{\rm max} = 0.07$, whereas the second oscillator corresponds to $\lambda_{\rm max} = 0.10$. It is confirmed from the aforesaid values of $\lambda_{\rm max}$ that the isolated R\"ossler oscillators exhibit chaotic dynamics at these parameter values; however, the second oscillator has a higher value of $\lambda_{\rm max}$ than the first. If we can define the magnitude of $\lambda_{\rm max}$ as the `degree of chaos' of any chaotic system, then Fig.~\ref{fig:ross2} infers that there exist a predominant net information transfer from the oscillator with a higher degree of chaos to the other.
\begin{figure}[h]
	\centering
	\includegraphics[width=30cm,height=2.9cm, keepaspectratio]{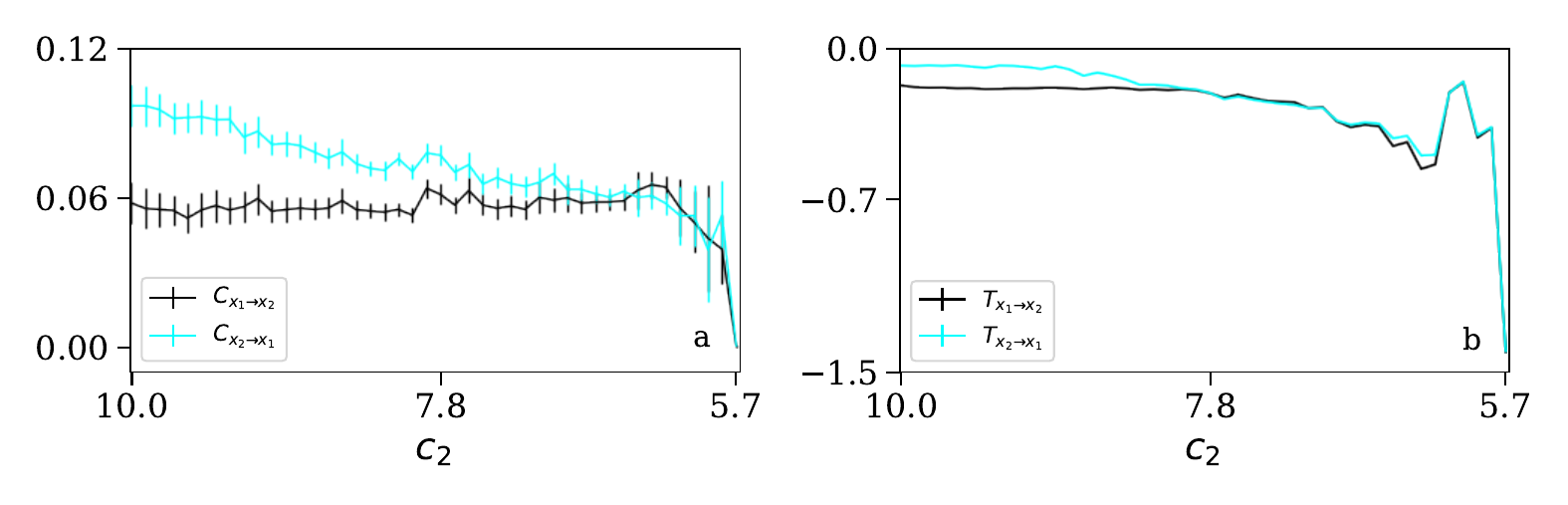}
	\caption{The CMI values and the LK information flow indices have been plotted as a function of $c_2$ in subplots (a) and (b), respectively. We have adopted other system parameters as $c_1 = 5.7$ and $\alpha =0.10$. The error bars in subplot~(b) are not visible because of their smaller values ($\sim 10^{-4}$).} 
	\label{fig:ross4}
\end{figure}
Lastly, we conduct a comprehensive analysis of CMI values and LK information flow indices as a function of the system parameter $c_2$ for coupled R\"ossler oscillators (Eq.~\ref{eq:ross}) with a constant coupling strength $\alpha = 0.10$ and $c_1 = 5.7$. The primary aim of this investigation is to study the information flow between the oscillators, particularly emphasizing the net transfer of information from the oscillator with the higher maximum Lyapunov exponent ($\lambda_{\rm max}$) to the other as a function of $c_2$. Figure~\ref{fig:ross4}a depicts the variation of CMI values as a function of $c_2$. The cyan plot measures the influence of $x_2(t)$ on $x_1(t+\tau)$, i.e., $C_{x_2 \rightarrow x_1}$. Besides, the black plot measures the influence of $x_1(t)$ on $x_2(t+\tau)$, i.e., $C_{x_1 \rightarrow x_2}$. Notably, the difference in CMI values diminishes as $c_2$ decreases monotonically, and these values converge when $c_2 = c_1 = 5.7$. A similar conclusion can also be made from the variation of $T_{x_1 \rightarrow x_2}$ and $T_{x_2 \rightarrow x_1}$ as a function of $c_2$, as shown in Fig.~\ref{fig:ross4}b.
Our analysis, therefore, reveals a predominant net information transfer from the oscillator with a higher degree of chaos to the other. In order to establish the robustness of these results, we have employed two different asymmetric indices: CMI and LK information flow. Given that both indices lead to similar conclusions, we primarily utilize the CMI index to illustrate the subsequent results in this paper.
\subsection{COM with different functional forms}
\label{sec:non_iden}
\begin{figure}[h]
	\centering
	\includegraphics[width=30cm,height=2.7cm, keepaspectratio]{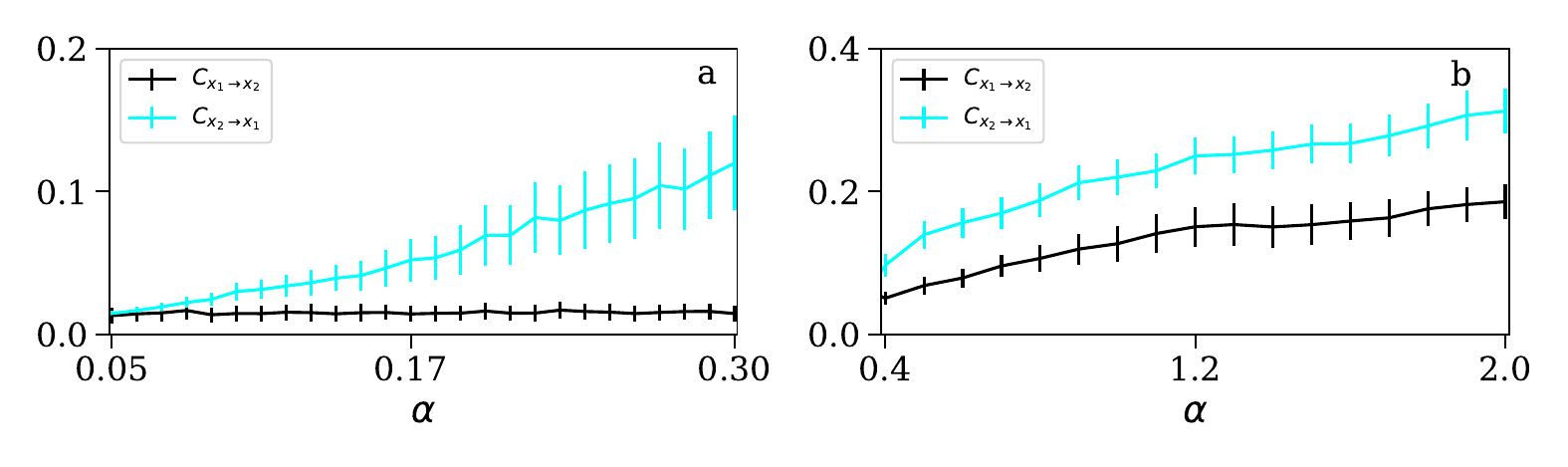}
	\caption{The CMI values have been plotted as a function of $\alpha$ for coupled R\"ossler-Lorenz oscillators (subplot~a) and for coupled Lorenz-Chen oscillators (subplot~b).} 
	\label{fig:rl1}
\end{figure}

Now, we switch to the second category, where the interacting oscillators have different functional forms. The general form of equations of motion is as follows:
\begin{subequations}
	\label{eq:non_iden}
	\begin{eqnarray}
		\frac{d{\mathbf{x}}_1}{dt} &=& \mathbf{F(x}_1) + \alpha \cdot \mathsf{C} \cdot (\mathbf{x}_2 - \mathbf{x}_1),\label{eq:non_iden_a}\\
		\frac{d{\mathbf{x}}_2}{dt} & =& \mathbf{G(x}_2) + \alpha \cdot \mathsf{C} \cdot  (\mathbf{x}_1 - \mathbf{x}_2),\label{eq:non_iden_b}
	\end{eqnarray}
\end{subequations}
where $\mathbf{F}(\cdot)$ and $\mathbf{G}(\cdot)$ are the mathematical forms of the interacting oscillators. Our primary objective is to ascertain the applicability of the conclusion outlined in Sec.~\ref{sec:para_mis} within the context of the current scenario. To this end, we adopt the example involving mutually coupled R\"ossler and Lorenz~\cite{lorenz63} oscillators. In the case of coupled R\"ossler and Lorenz oscillators, following Eq.~\ref{eq:non_iden}, we have $ \mathbf{F(x}_1) = [-y_1-z_1, x_1 + 0.2 y_1, 0.2 + (x_1 - 5.7)z_1]$, $ \mathbf{G(x}_2) = [10(y_2-x_2), x_2(28-z_2)-y_2, x_2 y_2 - \frac{8}{3}z_2]$, and $\mathsf{C} = \text{diag}(1, 0, 0)$. In the absence of coupling (i.e., when $\alpha = 0$), the values of $\lambda_{\rm max}$ for R\"ossler and Lorenz oscillators are $0.07$ and $0.90$, respectively. Therefore, based on the conclusion drawn in Sec.~\ref{sec:para_mis}, a predominant net information transfer from the Lorenz oscillator to R\"ossler oscillator is anticipated. Figure~\ref{fig:rl1}a illustrates the dependence of the CMI values on the coupling strength $\alpha$, revealing predominantly positive values for $(C_{x_2 \rightarrow x_1} - C_{x_1 \rightarrow x_2})$ across its range. 
Finally, we extend our study to the second example of this section: coupled Lorenz-Chen oscillators. In the context of coupled Lorenz and Chen~\cite{chen99} oscillators, in accordance with Eq.~\ref{eq:non_iden}, we establish the following mappings: $\mathbf{F(x}_1) = [10(y_1-x_1), x_1(28-z_1)-y_1, x_1 y_1 - \frac{8}{3}z_1]$, $\mathbf{G(x}_2) = [35(y_2-x_2), -7x_2-x_2 z_2+28z_2, x_2 y_2 - 3z_2]$. The coupling matrix $\mathsf{C}$ is represented as $\text{diag}(0, 0, 1)$. In the absence of coupling, the maximum Lyapunov exponent ($\lambda_{\rm max}$) values for Lorenz and Chen oscillators stand at $0.90$ and $2.02$, respectively~\cite{sprott03}. Consequently, in alignment with the inference of preceding examples, we observe a net information transfer from the Chen oscillator to the Lorenz oscillator (Fig.~\ref{fig:rl1}b). In both scenarios, the function $\mathbf{F(x}_1)$ is attributed to the oscillator exhibiting a lesser degree of chaos, while $\mathbf{G(x}_2)$ corresponds to the oscillator with a higher degree of chaos. Consequently, we ascertain a predominant net information transfer from $\mathbf{G(x}_2)$ to $\mathbf{F(x}_1)$ in both cases.
Until now, our analysis has been confined to a specific scenario where $\mathbf{F(x}_1)$ and $\mathbf{G(x}_2)$ exhibit identical dimensions in phase space. In Sec.~\ref{sec:diff_topo}, we broaden the scope of our investigation by transitioning our study to a more generalized framework.
\subsection{Coupled oscillators with different dimensions}
\label{sec:diff_topo}
It is important to emphasize that CMI is a model-free index. Consequently, our analysis retains its applicability even when the interacting oscillators possess different dimensions in phase space. With this underlying motivation, we have incorporated the 4D R\"ossler oscillator~\cite{ross79} into our study. The 4D R\"ossler oscillator is a four-dimensional, autonomous oscillator characterized by the following equations of motion:
\begin{subequations}
	\label{eq:hr}
	\begin{eqnarray}
		\frac{d{x}}{dt} &=& -y-z,\label{eq:hr_a}\\
		\frac{d{y}}{dt} &=& x+0.25y+w,\label{eq:hr_b}\\
		\frac{d{z}}{dt} &=& 3+xz,\label{eq:hr_c}\\
		\frac{d{w}}{dt} &=& -0.5z+0.05w.\label{eq:hr_d}
	\end{eqnarray}
\end{subequations}
The maximum Lyapunov exponent $\lambda_{\rm max}$ of this oscillator is $0.11$. Note that the Lyapunov spectrum of Eq.~\ref{eq:hr} comprises four distinct values, with two being positive and the other being negative. Consequently, Eq.~\ref{eq:hr} characterizes a hyperchaotic system.
\begin{figure}[htbp!]
	\includegraphics[width=30cm,height=2.8cm, keepaspectratio]{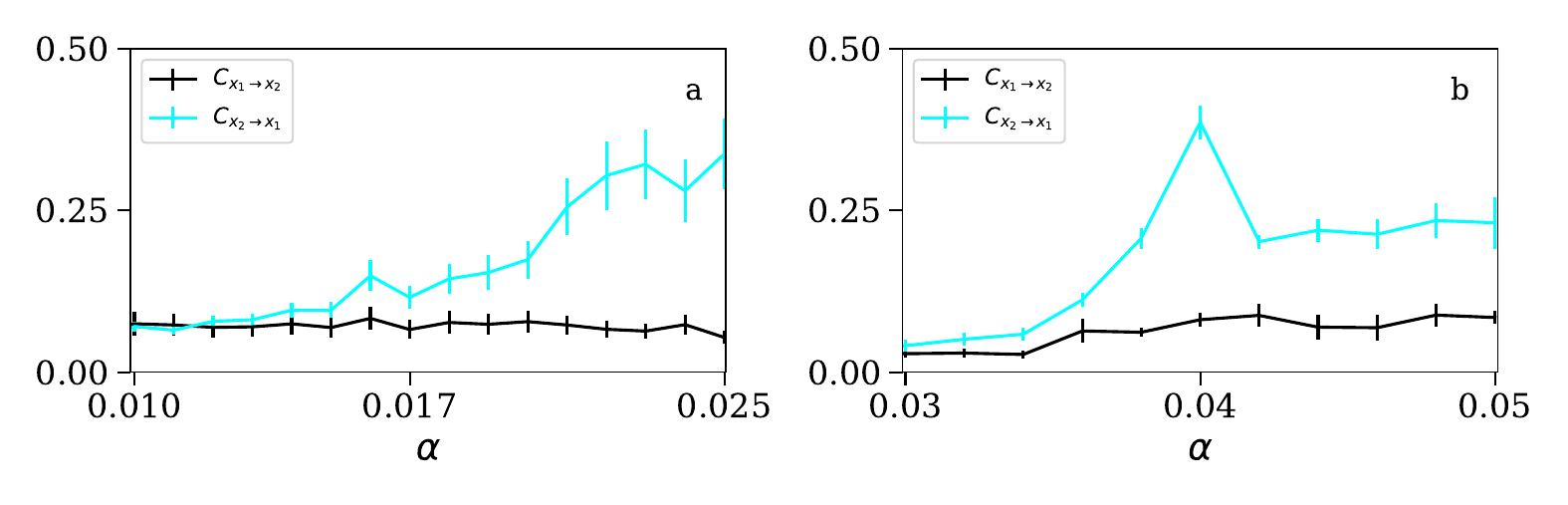}
	\caption{ The CMI values have been systematically calculated for two different configurations: coupled  R\"ossler-4D R\"ossler oscillators, as presented in subplot (a), and coupled 4D R\"ossler-Lorenz oscillators, depicted in subplot (b). These results are presented as functions of the coupling strength $\alpha$. }
	\label{fig:hr}
\end{figure}
In this analysis, we consider two distinct combinations. In the first combination, $\mathbf{F(x}_1) = [-y_1-z_1, x_1+ 0.2y_1, 0.2 + z_1(x_1-5.7)]$ represents the R\"ossler oscillator, while $\mathbf{G(x}_2)$ corresponds to the 4D R\"ossler oscillator (Eq.~\ref{eq:hr}). In contrast, $\mathbf{F(x}_1)$ denotes the 4D R\"ossler oscillator for the second combination, and $\mathbf{G(x}_2) = [10(y_2-x_2), x_2(28-z_2)-y_2, x_2 y_2 - \frac{8}{3}z_2]$ corresponds to the Lorenz oscillator~\cite{lorenz63}. The maximum Lyapunov exponents ($\lambda_{\rm max}$) of the Lorenz oscillator with $\alpha = 0$ is $0.90$. The key commonality between these two combinations lies in the fact that when the coupling is inactive (i.e., $\alpha = 0$), the oscillators associated with $\mathbf{F(x}_1)$ exhibit lower values $\lambda_{\rm max}$ in comparison to those linked with $\mathbf{G(x}_2)$. Consequently, it is reasonable to anticipate a net information transfer from $\mathbf{G(x}_2)$ to $\mathbf{F(x}_1)$. Finally, we calculate the CMI values as a function of coupling strength $\alpha$ for the two discussed models (Fig.~\ref{fig:hr}). Similar to the previous examples, here also, we observe a similar pattern in the variations of CMIs as a function of $\alpha$ for both combinations. 
\section{Conclusion}
\label{sec:conclusion}

In coupled oscillator models with unidirectional coupling, identifying the driver and driven relationship is pretty straightforward: information is transferred from the driving oscillator to the driven oscillator. This driver--driven concept, however, does not apply in the case of mutual (or bidirectional) coupling. This paper specifically addresses this issue. More explicitly, in this paper, we have focused on one of the fundamental questions in a coupled chaotic oscillator framework: \emph{What is the direction of net information transfer in mutually coupled non-identical oscillators?} To address this, we have adopted a particular form of CMI, an asymmetric and model-free index, to quantify the amount of information transferred from one oscillator to the other. Furthermore, we have defined the maximum Lyapunov exponent ($\lambda_{\rm max}$) as a quantitative measure of the `degree of chaos' in an isolated chaotic system. Our findings demonstrate a consistent trend: \emph{a pronounced predominant net information transfer from the oscillator exhibiting a higher degree of chaos to the other}. We have ascertained the broad applicability of our findings by considering two categories of coupled `non-identical' oscillators: In the first category, the functional forms of both oscillators are identical, with a disparity observed in one or more parameter values. An additional measure, the LK information flow, have been incorporated to support the generality of our results. In the second category, the functional forms of the oscillators are entirely distinct from one another.We have further expanded our investigation by examining the effectiveness of CMI to find the net information direction in coupled oscillator models, where the interacting oscillators exhibit different dimensions in phase space. These thorough analyses enable us to draw a conclusion that remains valid under various conditions, ensuring the robustness of our results.
\section*{Acknowledgment}
This study is supported by the Czech Academy of Sciences, Praemium Academiae
awarded to M. Palu\v{s}.
\bibliography{Ghosh_Palus_Manuscript.bib}

\end{document}